\documentclass{iau} 
\usepackage{graphicx}

\title[IAUS 316.~~Stellar population properties of massive GCs and UCDs in Fornax] %% give here short title %%
{Stellar population properties of the most massive globular clusters and 
ultra-compact dwarf galaxies of the Fornax cluster}

\author[Michael Hilker]   %% give here short author list %%
{Michael Hilker$^1$}
\affiliation{$^1$European Southern Observatory, Karl-Schwarzschild-Str.\,2, \\ 
D-85748, Garching bei M\"unchen, Germany \\ email: {\tt mhilker@eso.org}}
\pubyear{2015}
\volume{316}  %% insert here IAU Symposium No.
\jname{Formation, evolution, and survival of massive star clusters}
\editors{C. Charbonnel \& A. Nota, eds.}
\begin{document}

\maketitle

\begin{abstract}

Most ultra-compact dwarf galaxies (UCDs) and very massive globular clusters
reside in nearby galaxy clusters or around nearby giant galaxies. Due to their
distance ($>$4\,Mpc) and compactness ($r_{\rm eff}<100$\,pc) they are
barely resolved, and thus it is difficult to obtain their internal properties.
Here I present our most recent attempts to constrain the mass function,
stellar content and dynamical state of UCDs in the Fornax cluster.
Thanks to radial velocity membership assignment of $\sim$950 globular
clusters (GCs) and UCDs in the core of Fornax, the shape of their mass
function is well constrained. It is consistent with the `standard' Gaussian
mass function of GCs. Our recent simulations on the disruption process
of nucleated dwarf galaxies in cluster environments showed that
$\sim$40\% of the most massive UCDs should originate from nuclear star
clusters. Some Fornax UCDs actually show evidence for this scenario, as
revealed by extended low surface brightness disks around them and onsets
of tidal tails. Multi-band UV to optical imaging as well as low to medium
resolution spectroscopy revealed that there exist UCDs with youngish ages,
(sub-)solar [$\alpha$/Fe] abundances, and probably He-enriched populations.
\end{abstract}

\firstsection % if your document starts with a section,
\section{Introduction}

Most of the massive globular clusters (GCs) of our Milky Way show evidence
of multiple stellar populations with differences in their light element
abundances (see contributions of Piotto and Milone in this volume). A few
GCs even exhibit spreads in iron abundance and probably age. Those are
nuclear star cluster candidates whose host galaxies were disrupted during
the assembly history of the Milky Way.

\begin{figure}[t]
% \vspace*{-2.0 cm}
\begin{center}
\includegraphics[width=12.5cm]{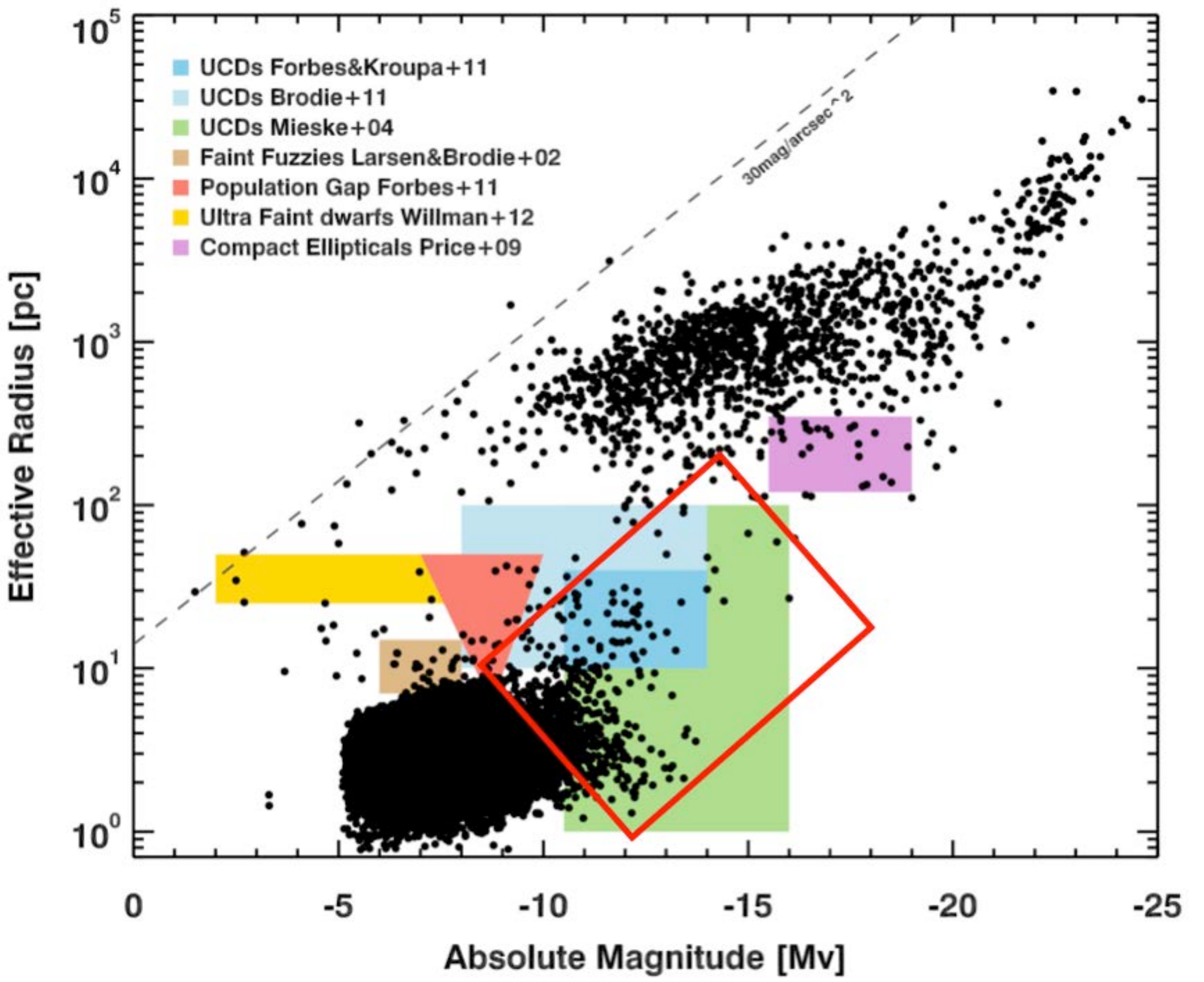} 
% \vspace*{-1.0 cm}
\caption{Luminosity-size plane of early-type stellar systems. The
compilation is mainly based on similar samples shown in \cite[Misgeld
\& Hilker (2011)]{Misgeld11} and \cite[Norris et al. (2014)]{Norris14}.
Coloured areas show different definitions of compact  objects.
Their references are given in the legend. The red rectangle encompasses
the objects of interest for this contribution. The dashed line 
indicates the surface brightness limit of 30\,mag/arcsec$^2$.}
\label{fig1}
\end{center}
\end{figure}

In galaxy clusters, disruption of low mass, nucleated galaxies was very 
common in the past. Indeed, in the Virgo and Fornax clusters there exists a 
large population of very massive and compact star cluster-like objects, 
called ultra-compact dwarf galaxies (UCDs). It was early
speculated that they might be either ``{\it very bright GCs}'' or lower mass
examples of ``{\it compact ellipticals like M\,32}'', or they might ``{\it
represent the nuclei of disolved dE,Ns}'' (\cite[Hilker et al. 1999]{Hilker99}).
In the recent years, more and more UCDs have been discovered, some very
compact, some rather diffuse, filling up the previously unpopulated region
between star clusters and dwarf galaxies in the luminosity-size plane (see
figure\,\ref{fig1}). Different authors apply different definitions for UCDs,
either invoking a lower luminosity/mass cut or a lower size cut.
Since those definitions are not physically motivated, I consider as 'UCDs' all
objects in the red box shown in figure\,\ref{fig1}. In general, UCDs are
characterized by an upper envelope in the mass-size relation and enhanced
dynamical mass-to-light ratios (\cite[Mieske et al. 2008b]{Mieske08b}),
roughly occurring at $>2\times10^6M_{\odot}$.

If one postulates that the complex
GCs $\omega$\,Centauri and M54 in the Milky Way and G1 in Andromeda
are low-mass UCDs, one would expect that UCDs in general should also
have complex star formation and chemical enrichment histories. However,
due to the large distance (most are located beyond 4\,Mpc) and thus
unresolved nature of UCDs, multiple stellar populations in them are very
difficult to detect and quantify. Most quantities have to be derived from
images and spectra of their unresolved stellar populations. The determination
of spatially resolved properties of UCDs requires space-based imaging
and adaptive optics assisted spectroscopy.

\section{The most massive globular clusters in the Fornax cluster}

The Fornax cluster has a very well studied GC and UCD population.
GC counts from photometric surveys revealed that there exist
$\sim6,450\pm700$ GCs within 83\,kpc projected distance around
the central cluster galaxy NGC\,1399 (\cite[Dirsch et al. 2003]{Dirsch03}).
Within 300\,kpc of NGC\,1399 the GC number counts increase to
$\sim11,100\pm2,400$ GCs (\cite[Gregg et al. 2009]{Gregg09}, derived
from the data of \cite[Bassino et al. 2006]{Bassino06}).
The number of radial velocity confirmed GCs and UCDs around NGC\,1399
is $\sim$950 (\cite[Schuberth et al. 2010]{Schuberth10}, Hilker, in prep.).
Within 300\,kpc of NGC\,1399 the sample of confirmed UCDs with masses
above $5\times10^6M_{\odot}$ is rather complete and the photometrically
selected GCs are well sampled below that limit. Thus, a mass function of
UCDs and GCs can be constructed with high confidence (see also
contribution of Schulz in this volume). The shape of that mass function
was already presented in \cite{Hilker09}, where we showed that the mass
function above $3\times10^5M_{\odot}$ can be approximated with two
power laws with slope exponents of $-1.9$ and $-2.7$ for objects
below and above $2\times10^6M_{\odot}$, respectively. This shape is
broadly consistent with the `standard' (Gaussian) mass function of GCs.
Thus, from this point of view, UCDs seem to be consistent with being
drawn from the GC mass function (\cite[Mieske et al. 2012]{Mieske12}).
However, the most
massive UCD in Fornax, UCD3, exhibits a compact core and an extended
stellar envelope of $\sim$90\,pc size (\cite[Evstigneeva et al.
2008]{Evstigneeva08}), pointing rather to a remnant of a disrupted,
nucleated dwarf galaxy.
 
\begin{figure}[t]
% \vspace*{-2.0 cm}
\begin{center}
\includegraphics[width=13.5cm]{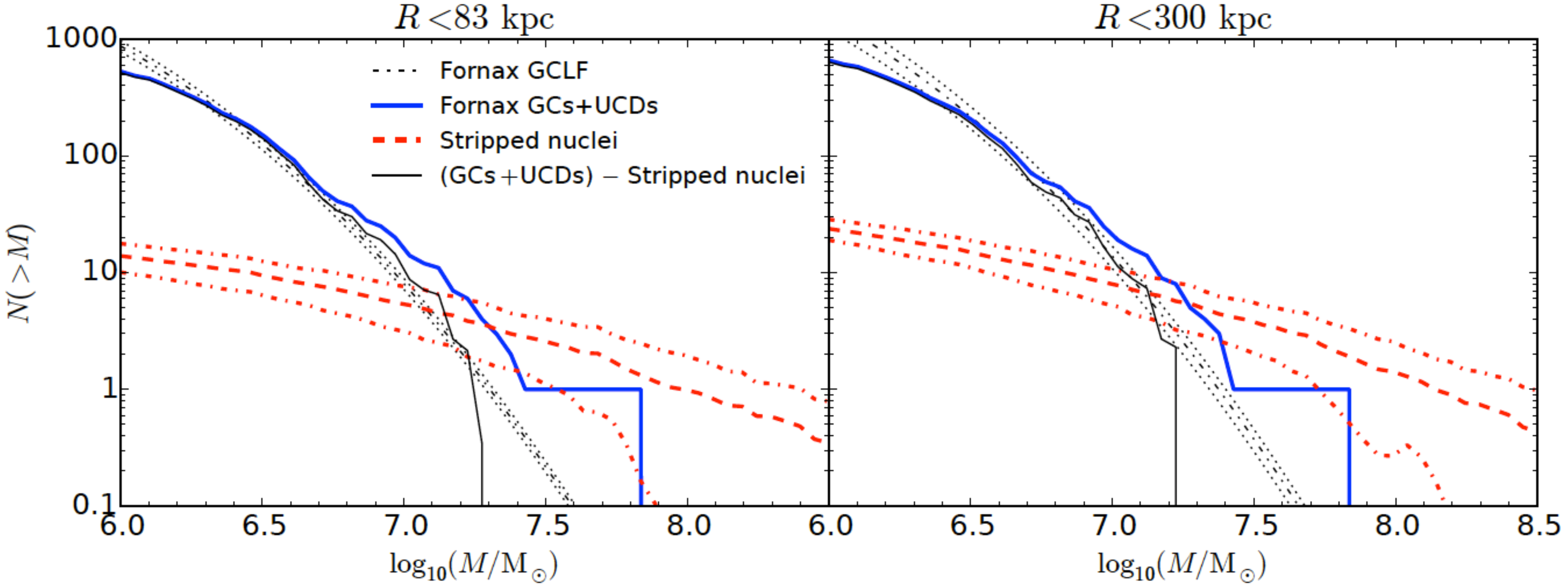} 
% \vspace*{-1.0 cm}
\caption{Cumulative mass function of simulated stripped nuclei in
Fornax-like clusters within a projected distance of 83 (left panel)
and 300 kpc (right panel) from the central galaxy (dashed line with
the standard deviation between clusters shown by dash-dotted lines)
compared with GCs/UCDs in the Fornax cluster (thick solid line).
The thin dash-dotted line shows the integrated Fornax GC luminosity
function (GCLF) with the standard deviation given by dotted lines.
The thin solid line shows the mean for the simulations subtracted
from the Fornax GCLF. Figure taken from \cite{Pfeffer15}.}
\label{fig2}
\end{center}
\end{figure}

We asked ourselves how many UCDs that originated from nuclei of
disrupted galaxies are expected to contribute to the mass function in a
Fornax cluster-like environment? We conducted cosmological
simulations combined with a semi-analytic galaxy formation model and
empirical prescriptions for the nucleation fraction and nuclei to galaxy
mass fraction to identify stripped nuclei of disrupted satellite galaxies
(\cite[Pfeffer et al. 2014]{Pfeffer14}). In figure\,\ref{fig2} we show the 
cumulative mass function of the simulated stripped nuclei and compare
it with the mass function of observed GCs/UCDs with masses
$>$10$^6M_{\odot}$. As one can see, stripped nuclei can only explain
the number of very massive UCDs. For masses larger than $10^7M_{\odot}$
we predict stripped nuclei account for $\sim$40\% of GCs/UCDs. For
masses between $10^6$ and $10^7M_{\odot}$ the stripped nuclei
account for only $\sim$2.5\% of GCs/UCDs. Thus, the majority of lower
mass UCDs should be of star cluster origin, either formed as very
massive genuine globular cluster or being the result of merged super
star cluster complexes (e.g., \cite[Fellhauer \& Kroupa
2002]{Fellhauer02}, \cite[Br\"uns et al. 2009]{Bruens09}).

Indeed, there exist young massive star clusters that occupy the
same space in the mass-size plane as UCDs (\cite[Kissler-Patig et al.
2006]{Kissler06}). Thus, mass and size alone are not indicative of
the origin of UCDs. 

\section{Internal properties of ultra-compact dwarf galaxies}

The nature of UCDs can be revealed, or at least constrained,
by analysing the spatially resolved surface brightness profiles and
internal kinematics and by studying the properties of their unresolved stellar
population. A supermassive black hole (SMBH) and/or and extended
star formation history would favour their origin as stripped nuclei of
a dwarf galaxy. A high metallicity, i.e. not obeying the mass-metallicity
relation of nuclei, would rather point to a star cluster
origin in an enriched (maybe merger-induced) environment.

\cite{Evstigneeva08} revealed positive colour gradients in the
resolved light profiles of some UCDs in Fornax and Virgo. A blue
nucleus and a red envelope are consistent with the colour differences
of nuclei and stellar body in nucleated dwarf ellipticals (\cite[Lotz et
al. 2004]{Lotz04}), where nuclei are formed from merging, metal-poor
GCs.

In a recent study of the structural composition and clustering
properties of 97 UCDs in the halo of NGC\,1399, we found evidence
for faint stellar envelopes around several UCDs with effective radii of
up to 90\,pc (\cite[Voggel et al. 2015]{Voggel15}). One particularly
extended UCD shows clear signs of tidal tails extending out to
$\sim$350\,pc.
We also detected, in a statistical sense, a local overdensity of blue
GCs on scales of $\leq1$\,kpc around UCDs.
These could either be remnant GCs of a formerly rich GC system
around a disrupted dwarf galaxy, or surviving star clusters
of a merged super star cluster complex (e.g., \cite[Br\"uns et al.
2009]{Bruens09}).

Combining GALEX NUV/FUV with optical photometry,
\cite{Mieske08a} showed that the location of several Fornax UCDs
and massive GCs in the (FUV-NUV) versus (NUV-V) plane can only
be explained with He-enhanced SSP models. This might be a hint to
multiple populations in these GCs/UCDs, since He enhancement is
also observed in the second generation stars of multi-population
Galactic GCs.

\begin{figure}[t]
% \vspace*{-2.0 cm}
\begin{center}
\includegraphics[width=12.5cm]{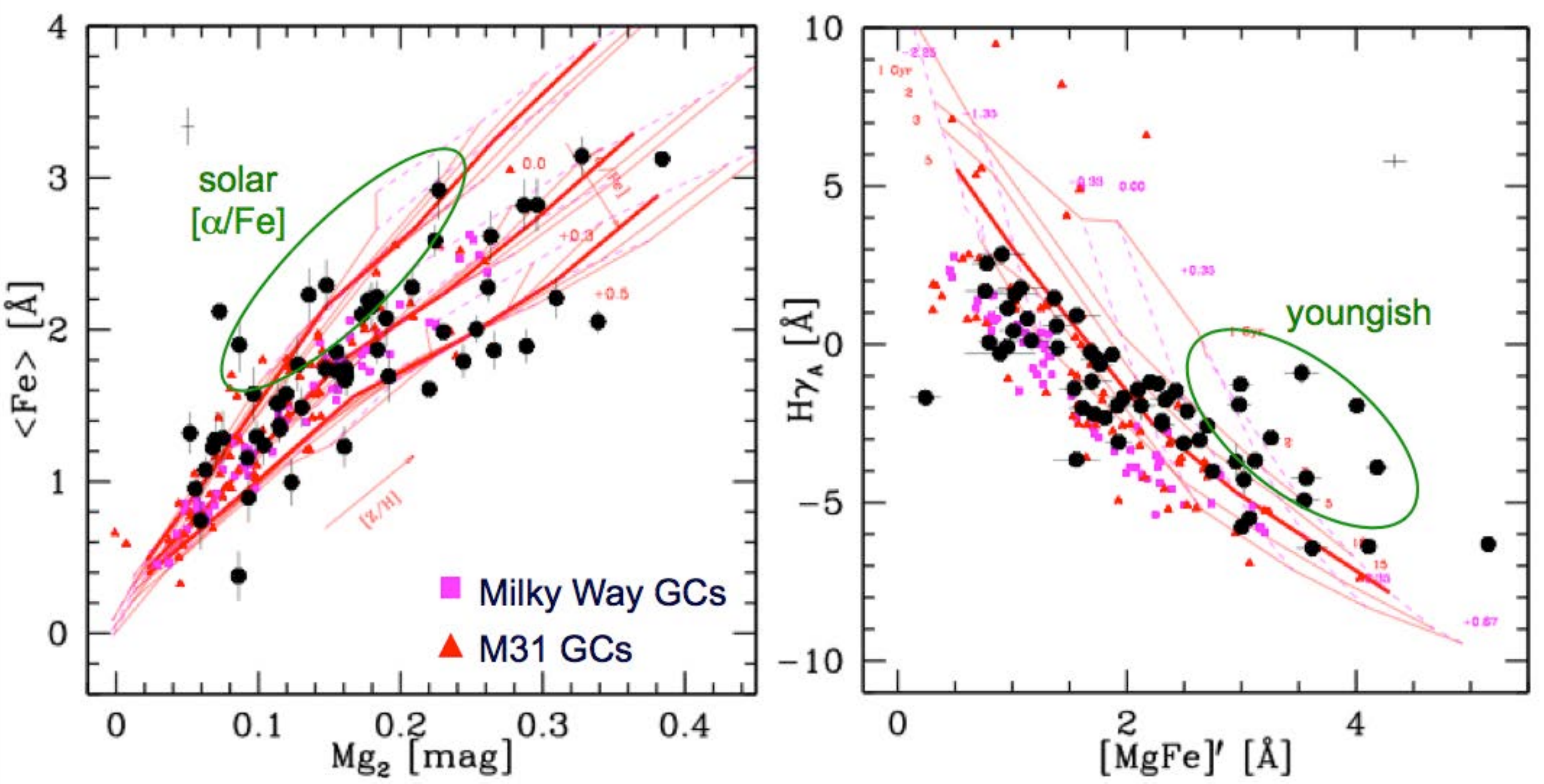} 
% \vspace*{-1.0 cm}
\caption{Lick index measurements of $\sim$50 UCDs and massive GCs
in the core of the Fornax cluster (black dots) compared to Milky Way and
M31 GCs (as indicated). {\it Left panel}: Mg$_2$ versus iron index, which
allows to differentiate [$\alpha$/Fe] abundances. {\it Right panel}: 
Metallicity index [MgFe]' versus the Balmer index H$_{\gamma_A}$, which
breaks the age-metallicity degeneracy. The model grids are based on 
SSP models from \cite{Thomas03}. Objects with (sub-)solar [$\alpha$/Fe]
values and young ages are encircled by green ellipses.}
\label{fig3}
\end{center}
\end{figure}

Deep VLT/FORS2 spectroscopy on $\sim$50 bright GCs ($M_V<-9$
mag) allowed us to measure Lick indices (Hilker, Puzia, et al., in
prep.). In Fig.\,\ref{fig3} we show two planes of Lick indices
that constrain the [$\alpha$/Fe] abundances and break the
age-metallicity degeneracy, respectively. Besides old GCs with 
enhanced [$\alpha$/Fe] abundances, there also exist objects with
(sub-solar) [$\alpha$/Fe] values and some metal-rich
intermediate-age GCs/UCDs (2-7 Gyr). Those might
have had their origin in dwarf galaxies with low star formation rates
and/or as nuclear star clusters with recurrent star formation episodes.

\begin{figure}[t]
% \vspace*{-2.0 cm}
\begin{center}
\includegraphics[width=11.0cm]{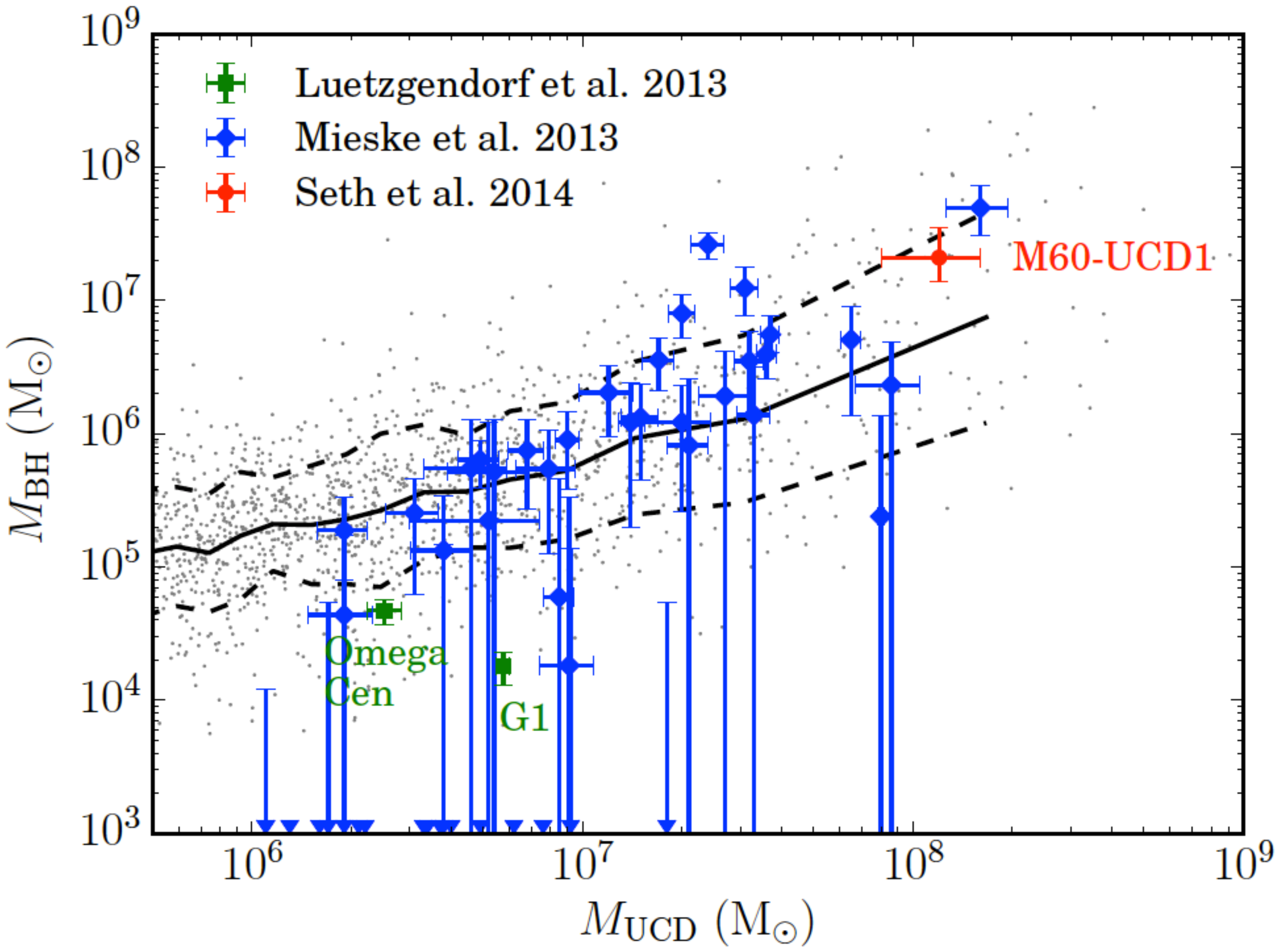} 
% \vspace*{-1.0 cm}
\caption{Predicted masses of central black holes in stripped nuclei
as function of stripped nucleus mass. The mean and 1-$\sigma$
confidence interval for the simulations (grey dots) are given by the
solid and dashed lines, respectively, using bin sizes of 100 objects.
The typical 1-$\sigma$ confidence interval in MBH is 0.5 dex.
For comparison, we also show the black hole mass of M60-UCD1
(\cite[Seth et al. 2014]{Seth14}), the inferred black hole masses of
UCDs assuming elevated mass-to-light ratios are due to central
black holes (\cite[Mieske et al. 2013]{Mieske13}) and the limits for
central black holes in the GCs $\omega$\,Cen and G1
(\cite[L\"utzgendorf et al. 2013]{Nora13}). For the observed UCDs
and GCs, objects with implied black hole masses of zero are given
by triangles at the bottom of the figure. Figure taken from
\cite{Pfeffer15}.}
\label{fig4}
\end{center}
\end{figure}

The most convincing case of a stripped nuclei origin of a UCD is the
discovery of a SMBH in M60-UCD1, one of the most massive and
densest UCDs in the Virgo cluster (\cite[Seth et al. 2014]{Seth14}).
The SMBH comprises 15\% of the UCD's total mass. In our simulations
of stripped nuclei (\cite[Pfeffer et al. 2015]{Pfeffer15}) we predict the
masses of their central black holes (BHs) as function of stripped nuclei
mass (see figure\,\ref{fig4}). The SMBH mass of M60-UCD1 falls on
top of the simulations, and, interestingly, also the inferred BH masses
of UCDs with elevated mass-to-light ratios (\cite[Mieske et al.
2013]{Mieske13}) are very much consistent with the simulations.
More AO-assisted spectroscopic observations are underway to prove
or disprove the SMBH hypothesis
for UCDs with high dynamical mass-to-light ratios.

\section{Summary and outlook}

Our general conclusions from the findings on UCDs and massive GCs
in the Fornax cluster presented in this contribution can be summarized
as follows:

\begin{itemize}
\item UCDs are defined through an upper envelope in the mass-size
relation and enhanced dynamical mass-to-light ratios –- {\it roughly
occurring at $>2\times10^6M_{\odot}$}.
\item UCDs share properties of nuclei as well
as young massive star clusters, e.g. the mass-size relation. 
{\it They are a mixed bag of objects.}
\item The mass function of massive GCs/UCDs in Fornax shows a break
at $\sim2\times10^6M_{\odot}$. There is no excess of UCD-mass
objects. {\it Thus, they are mostly of `star cluster origin‘.}
\item Stripping of nucleated galaxies in a cosmological framework
cannot explain the total number of UCDs, {\it but $\sim$40\% of UCDs
with a mass $>10^7M_{\odot}$ are compatible with stripped nuclei.}
\item M60-UCD1 has a SMBH, which makes
up 15\% of the UCD’s mass. {\it This is direct evidence for the stripping
scenario as valid channel to form UCDs.}
\item Extended stellar envelopes and overdensities of star clusters 
around UCDs might hint at the accretion of nucleated dwarf galaxies
or at a dissolving super star clusters that were formed from merged
star cluster complexes.
\end{itemize}

Distinguishing stripped nuclei from super star clusters or massive GCs
is a difficult task. Many internal properties of UCDs can be explained
by both scenarios. More measurements of SMBHs in UCDs, derivation
of their star formation histories and light element abundances are
needed to unveil the nature of individual UCDs, and thus build up
statistically meaningful samples for UCDs of different flavours.


\begin{thebibliography}{}

\bibitem[Bassino et al. (2006)]{Bassino06}
{Bassino, L.P., Faifer, F.R., Forte, J.C., Dirsch, B., Richtler, T., Geisler, D.,
\& Schuberth, Y.} 2006,
\textit{A\&A}, 451, 789

\bibitem[Br\"uns et al. (2009)]{Bruens09}
{Br\"uns, R.C., Kroupa, P., \& Fellhauer, M.} 2009,
\textit{ApJ}, 702, 1268

\bibitem[Brodie et al. (2011)]{Brodie11}
{Brodie, J.P., Romanowsky, A.J., Strader, J., \& Forbes, D.A.} 2011,
\textit{AJ}, 142, 199

\bibitem[Dirsch et al. (2003)]{Dirsch03}
{Dirsch, B., Richtler, T., Geisler, D., Forte, J.C., Bassino, L.P., \& Gieren,
W.P.} 2003,
\textit{AJ}, 125, 1908

\bibitem[Evstigneeva et al. (2008)]{Evstigneeva08}
{Evstigneeva, E.A., Drinkwater, M.J., Peng, C.Y., Hilker, M.,
De Propris, R., et al.} 2008
\textit{AJ}, 136, 461

\bibitem[Fellhauer \& Kroupa (2002)]{Fellhauer02}
{Fellhauer, M., \& Kroupa, P.} 2002,
\textit{MNRAS}, 330, 642

\bibitem[Forbes \& Kroupa (2011)]{ForbesK11}
{Forbes, D.A., \& Kroupa, P.} 2011
\textit{PASA}, 28, 77

\bibitem[Forbes et al. (2013)]{Forbes13}
{Forbes, D.A., Pota, V., Usher, C., Starder, J., Romanowsky, A.J., et al.}
2013
\textit{MNRAS}, 435, L6

\bibitem[Gregg et al. (2009)]{Gregg09}
{Gregg, M.D., Drinkwater, M.J., Evstigneeva, E., Jurek, R., Karick, A.M.,
Phillipps, S., Bridges, T., Jones, J.B., Bekki, K., \& Couch, W.J.} 2009
\textit{AJ}, 137, 498

\bibitem[Hilker et al. (1999)]{Hilker99}
{Hilker, M., Infante, L., Vieira, G., Kissler-Patig, M., \& Richtler, T.}
1999,
\textit{A\&AS}, 134, 75

\bibitem[Hilker (2009)]{Hilker09}
{Hilker, M.} 2009, in: S. Roeser (ed.),
\textit{Reviews in Modern Astronomy: Formation and Evolution of
Cosmic Structures}, Vol.\,21, (Wiley-VCH Verlag GmbH \& Co. KGaA),
p.\,199

\bibitem[Kissler-Patig et al. (2006)]{Kissler06}
{Kissler-Patig, M., Jord\'an, A., \& Bastian, N.} 2006
\textit{A\&A}, 448, 1031

\bibitem[Larsen \& Brodie (2002)]{Larsen02}
{Larsen, S.S., \& Brodie, J.P.} 2002
\textit{AJ}, 123, 1488

\bibitem[Lotz et al. (2004)]{Lotz04}
{Lotz, J.M., Miller, B.W., \& Ferguson, H.C.}
\textit{ApJ}, 613, 262

\bibitem[L\"utzgendorf et al. (2013)]{Nora13}
{L\"utzgendorf, N., Kissler-Patig, M., Neumayer, N., Baumgardt, H.,
Noyola, E., et al.} 2013 
\textit{A\&A}, 555,26

\bibitem[Mieske et al. (2004)]{Mieske04}
{Mieske, S., Hilker,M., \& Infante, L.} 2004
\textit{A\&A}, 418, 445

\bibitem[Mieske et al. (2008)]{Mieske08a}
{Mieske, S., Hilker,M., Bomans, D.J., Rey, S.-C., Kim, S., Yoon, S.-J.,
\& Chung, C.} 2008
\textit{A\&A}, 489, 1023

\bibitem[Mieske et al. (2008)]{Mieske08b}
{Mieske, S., Hilker,M., Jord\'an, A., Infante, L., Kissler-Patig, M., et al.}
2008
\textit{A\&A}, 487, 921

\bibitem[Mieske et al. (2012)]{Mieske12}
{Mieske, S., Hilker, M., \& Misgeld, I.} 2012
\textit{A\&A}, 537, 3

\bibitem[Mieske et al. (2013)]{Mieske13}
{Mieske, S., Frank, M.J., Baumgardt, H., L\"utzgendorf, N.,
Neumayer, N., \& Hilker, M.} 2013
\textit{A\&A}, 558, 14

\bibitem[Misgeld et al. (2011)]{Misgeld11}
{Misgeld, I., \& Hilker, M.} 2011,
\textit{MNRAS}, 414, 3699

\bibitem[Norris et al. (2014)]{Norris14}
{Norris, M.A., Kannappan, S.J., Forbes, D.A.; Romanowsky, A.J.,
Brodie, J.P., et al.} 2014
\textit{MNRAS}, 443, 1151

\bibitem[Pfeffer et al. 2014]{Pfeffer14}
{Pfeffer, J., Griffen, B.F., Baumgardt, H., \& Hilker, M.} 2014,
\textit{MNRAS}, 444, 3670

\bibitem[Pfeffer et al. (2015)]{Pfeffer15}
{Pfeffer, J., Hilker, M., \& Baumgardt, H.} 2015
\textit{MNRAS}, submitted

\bibitem[Schuberth et al. (2010)]{Schuberth10}
{Schuberth, Y., Richtler, T., Hilker, M., Dirsch, B., Bassino, L.P.,
Romanowsky, A.J., \& Infante, L.} 2010
\textit{A\&A}, 513, 52

\bibitem[Seth et al. (2014)]{Seth14}
{Seth, A.C., van den Bosch, R., Mieske, S., Baumgardt, H., Brok, M. Den,
et al.} 2014,
\textit{Nature}, 513, 398

\bibitem[Thomas et al. (2003)]{Thomas03}
{Thomas, D., Maraston, C. \& Bender, R.} 2003
\textit{MNRAS}, 339, 897

\bibitem[Voggel et al. (2015)]{Voggel15}
{Voggel, K., Hilker, M., \& Richtler, T.} 2015
\textit{A\&A}, accepted

\bibitem[Willman \& Strader (2012)]{Willman12}
{Willman, B., \& Strader, J.} 2012
\textit{AJ}, 144, 76

\end{thebibliography}
\end{document}